# Kinetic Equations for a Dirty Two-Band Superconductor

V.L.OSTROVSKI

ORT Braude College, Karmiel 21982, Israel

E-mail: vostrovs@braude.ac.il

**Abstract**

Green's function techniques for studying nonequilibrium processes in dirty two-band superconductors are discussed. Perturbation expansions and Green's function equations are developed. A time dependent modification of the Usadel equation is given.

## Introduction.

The time dependent Gorkov equation has proven to be very successful in the theory of superconductivity [1-11]. The usual description of superconductivity characterizes states by a one-component order parameter. However the properties of some compounds (e.g. $MgB_2, Nb_2Se, V_3Si$) are described by a two-band model [12] with a two- component order parameter. The two-gap model (the two gaps arise from the existence of two bands) was proposed [13, 14] at the dawn of superconductivity theory. The interest in the two-gap model has recently been renewed after the discovery of superconductivity in magnesium diboride[15-42]. The following study is based on a set of kinetic equations for the order parameters of a two-band system. We focus our attention mainly at the overlapping of energy bands on the Fermi surface, and effects of impurities. The theory of dirty two-band superconductivity has been developed within Matsubara formalism [15, 17, 19, 32, 43-46].

In this paper we develop the quasiclassical Keldysh-Usadel model for multi-band superconductors, following the real-time formalism [47-50].

## 1. Green's functions in the Nambu-Keldysh representation

The two-band pairing Hamiltonian in the BCS theory, ignoring impurity scattering, is given by

$$\hat{H} = \hat{H}_0 + \hat{H}_{\text{int}} \tag{1.1}$$

Where

$$\hat{H}_0 = \int d^3\vec{r} \sum_{n\sigma} \hat{\psi}^{+}_{n\sigma}(\vec{r}) h_n(\vec{r},t) \hat{\psi}_{n\sigma}(\vec{r}) \tag{1.2}$$

is the Hamiltonian of noninteracting quasiparticles, $h_n(\vec{r},t) = \frac{1}{2m_n}\left[-i\nabla_{\vec{r}} - e\vec{A}(\vec{r},t)\right]^2 + e\varphi(\vec{r},t) - \mu + \zeta_n$ , $\mu$ is the chemical potential , $n = 1,2$ are the band indexes, $m_n$ is the effective mass, $\zeta_n$ is the lowest energy level in the $n$-th band, $\vec{A},\varphi$ are potentials of external electromagnetic field and $\hat{\psi}^{+}_{n\sigma},\hat{\psi}_{n\sigma}$ are the creation and annihilation field operators in the Heisenberg representation for quasiparticles in band $n$ with spin $\sigma = \uparrow$ or $\downarrow$ .

$$\begin{aligned}
\hat{H}_{\text{int}} &= -\sum_n \int d^3\vec{r}\left[\Delta_n(\vec{r})\hat{\psi}^{+}_{n\downarrow}(\vec{r})\hat{\psi}^{+}_{n\uparrow}(\vec{r}) + \Delta^{*}_{n}(\vec{r})\hat{\psi}_{n\uparrow}(\vec{r})\hat{\psi}_{n\downarrow}(\vec{r})\ \right], \\
\Delta_n(\vec{r}) &= -\sum_m V_{nm}\left\langle \hat{\psi}_{m\downarrow}(\vec{r})\hat{\psi}_{m\uparrow}(\vec{r})\right\rangle, \\
\Delta^{*}_{n}(\vec{r}) &= -\sum_m V_{nm}\left\langle \hat{\psi}^{+}_{m\downarrow}(\vec{r})\hat{\psi}^{+}_{m\uparrow}(\vec{r})\right\rangle
\end{aligned} \tag{1.3}$$

is the Hamiltonian of pairing interaction, $\Delta_n(\vec{r},t)$ is the gap function of the n-th band , $V_{11},V_{22}$ are the constants of intraband interaction, $V_{12}, V_{21}$ are the constants of interband interaction $(V_{nm} = V_{mn})$.The model assumes the formation of Cooper pairs in each band and the tunneling of these pairs as a whole from one band into another. The equations of motion in the Heisenberg picture are

$$i\frac{\partial\hat{\psi}_{n\uparrow}(\vec{r},t)}{\partial t} = \left[\frac{1}{2m_n}\left(-i\nabla_{\vec{r}} - e\vec{A}(\vec{r},t)\right)^2 - \mu + \zeta_n\right]\hat{\psi}_{n\uparrow}(\vec{r},t) + \Delta_n(\vec{r},t)\hat{\psi}^+_{n\downarrow}(\vec{r},t) \qquad (1.4)$$

(for the spin-up component)

$$-i\frac{\partial\hat{\psi}^+_{n\downarrow}(\vec{r},t)}{\partial t} = \left[\frac{1}{2m_n}\left(i\nabla_{\vec{r}} - e\vec{A}(\vec{r},t)\right)^2 - \mu + \zeta_n\right]\hat{\psi}^+_{n\downarrow}(\vec{r},t) - \Delta^*_n(\vec{r},t)\hat{\psi}_{n\uparrow}(\vec{r},t) \qquad (1.5)$$

(for the spin-down component)

We introduce Nambu's pseudo-spinor field

$$\Psi_n(\vec{r},t) = \begin{pmatrix}\Psi_{n1}(\vec{r},t)\\ \Psi_{n2}(\vec{r},t)\end{pmatrix}, \qquad (1.6)$$
$$\Psi_{n1}(\vec{r},t) = \hat{\psi}_{n\uparrow}(\vec{r},t), \Psi_{n2}(\vec{r},t) = \hat{\psi}^+_{n\downarrow}(\vec{r},t)$$

with anti-commutation rules

$$\left\{\Psi_{n\alpha}(\vec{r},t), \Psi^+_{m\beta}(\vec{r}\,',t')\right\} \equiv \Psi_{n\alpha}(\vec{r},t)\Psi^+_{m\beta}(\vec{r}\,',t') + \Psi^+_{m\beta}(\vec{r}\,',t')\Psi_{n\alpha}(\vec{r},t) =$$
$$= \delta_{nm}\delta_{\alpha\beta}\delta(\vec{r}-\vec{r}\,')\delta(t-t'),$$
$$\left\{\Psi_{n\alpha}(\vec{r},t), \Psi_{m\beta}(\vec{r}\,',t')\right\} = \left\{\Psi^+_{n\alpha}(\vec{r},t), \Psi^+_{m\beta}(\vec{r}\,',t')\right\} = 0$$

The Hamiltonian of the system in Nambu's representation is

$$H=\sum_{n=1,2}\int d\vec{r}\,\Psi_n^{+}(\vec{r},t)\left(\hat{H}_0(\vec{r},t)+\hat{\Sigma}_n(\vec{r},t)\right)\Psi_n(\vec{r},t),$$
$$\hat{H}_0(\vec{r},t)=\begin{pmatrix} h_n(\vec{r},t) & 0 \\ 0 & -h^{*}_{n}(\vec{r},t)\end{pmatrix},\hat{\Sigma}_n(\vec{r},t)=\begin{pmatrix} 0 & \Delta_n(\vec{r},t) \\ \Delta^{*}_{n}(\vec{r},t) & 0\end{pmatrix} \quad (1.7)$$

The contour-ordered Green's function is defined by

$$\hat{G}_n(\vec{r},t;\vec{r}\,',t')=-i\left\langle T_{c_t}\left(\Psi_n(\vec{r},t)\Psi^{+}{}_{n}(\vec{r}\,',t')\right)\right\rangle \quad (1.8)$$

where $c_t$ is the Schwinger-Keldysh closed time path contour [49]

The greater Green's function is

$$\hat{G}_n^{>}(\vec{r},t;\vec{r}\,',t')=-i\left\langle\left(\Psi_n(\vec{r},t)\Psi^{+}{}_{n}(\vec{r}\,',t')\right)\right\rangle=$$
$$=-i\begin{pmatrix} \left\langle\psi_{n\uparrow}(\vec{r},t)\psi^{+}{}_{n\uparrow}(\vec{r}\,',t')\right\rangle & \left\langle\psi_{n\uparrow}(\vec{r},t)\psi_{n\downarrow}(\vec{r}\,',t')\right\rangle \\ \left\langle\psi^{+}{}_{n\downarrow}(\vec{r},t)\psi^{+}{}_{n\uparrow}(\vec{r}\,',t')\right\rangle & \left\langle\psi^{+}{}_{n\downarrow}(\vec{r},t)\psi_{n\downarrow}(\vec{r}\,',t')\right\rangle\end{pmatrix} \quad (1.9)$$

The lesser Green's function is then

$$\hat{G}_n^{<}(\vec{r},t;\vec{r}\,',t')=-i\left\langle\left(\Psi^{+}{}_{n}(\vec{r},t)\Psi_n(\vec{r}\,',t')\right)\right\rangle=$$
$$=i\begin{pmatrix} \left\langle\psi^{+}{}_{n\uparrow}(\vec{r}\,',t')\psi_{n\uparrow}(\vec{r},t)\right\rangle & \left\langle\psi_{n\downarrow}(\vec{r}\,',t')\psi_{n\uparrow}(\vec{r},t)\right\rangle \\ \left\langle\psi^{+}{}_{n\uparrow}(\vec{r}\,',t')\psi^{+}{}_{n\downarrow}(\vec{r},t)\right\rangle & \left\langle\psi_{n\downarrow}(\vec{r}\,',t')\psi^{+}{}_{n\downarrow}(\vec{r},t)\right\rangle\end{pmatrix} \quad (1.10)$$

We define the retarded and advanced Green's functions

$$\hat{G}_n^R(\vec{r},t;\vec{r}',t') = \theta(t-t')\left[\hat{G}_n^>(\vec{r},t;\vec{r}',t') - \hat{G}_n^<(\vec{r},t;\vec{r}',t')\right],$$
$$\hat{G}_n^A(\vec{r},t;\vec{r}',t') = -\theta(t'-t)\left[\hat{G}_n^>(\vec{r},t;\vec{r}',t') - \hat{G}_n^<(\vec{r},t;\vec{r}',t')\right], \qquad (1.11)$$

the Keldysh Green's function

$$\hat{G}_n^K(\vec{r},t;\vec{r}',t') = \hat{G}_n^>(\vec{r},t;\vec{r}',t') + \hat{G}_n^<(\vec{r},t;\vec{r}',t'), \qquad (1.12)$$

and the $4\times4$ triagonal matrix Green's function

$$G_n(\vec{r},t;\vec{r}',t') = \begin{pmatrix} \hat{G}_n^R(\vec{r},t;\vec{r}',t') & \hat{G}_n^K(\vec{r},t;\vec{r}',t') \\ 0 & \hat{G}_n^A(\vec{r},t;\vec{r}',t') \end{pmatrix}$$

The non-equilibrium Dyson equations for the matrix Green's function become

$$(G^{-1}{}_{0n}\otimes G_n)(\vec{r},t;\vec{r}',t') = \delta(\vec{r}-\vec{r}')\delta(t-t') + (\Sigma_n\otimes G_n)(\vec{r},t;\vec{r}',t')$$
$$(G_n\otimes G_{0n}^{-1})(\vec{r},t;\vec{r}',t') = \delta(\vec{r}-\vec{r}')\delta(t-t') + (G_n\otimes\Sigma_n)(\vec{r},t;\vec{r}',t') \qquad (1.13)$$

Where we have introduced the inverse free matrix Green's function

$$G^{-1}{}_{0n}(\vec{r},t;\vec{r}',t') = G^{-1}{}_{0n}(\vec{r},t)\delta(\vec{r}-\vec{r}')\delta(t-t'),$$
$$G^{-1}{}_{0n}(\vec{r},t) = i\tau^{(3)}\frac{\partial}{\partial t} + \frac{1}{2m_n}\left[\frac{\partial}{\partial\vec{r}} - ie\tau^{(3)}\vec{A}(\vec{r},t)\right]^2 - e\varphi(\vec{r},t) + \mu - \zeta_n, \qquad (1.14)$$
$$\tau^{(3)} = \begin{pmatrix} \tau^3 & 0 \\ 0 & \tau^3 \end{pmatrix}, \tau^3 = \begin{pmatrix} 1 & 0 \\ 0 & -1 \end{pmatrix}$$

The $4\times4$ matrix $\Sigma_n = \begin{pmatrix} \hat{\Sigma}_n^R & \hat{\Sigma}_n^K \\ 0 & \hat{\Sigma}_n^A \end{pmatrix}$ is the self-energy matrix. The operator $\otimes$ has the effect of integrating over the space and time coordinates and performing matrix multiplication when applied between two matrices:

$$\left[\left(A\otimes B\right)\right]_{ij}\left(\vec{r},t;\vec{r}',t'\right)=\int d\vec{r}''\int dt''\sum_k A_{ik}\left(\vec{r},t;\vec{r}'',t''\right)B_{kj}\left(\vec{r}'',t'';\vec{r}',t'\right)$$

Subtracting equations (1.13) we obtain the left-right Dyson equation

$$\left[G_{0n}^{-1}-\Sigma_n \overset{\otimes}{,} G_n\right]_- =0 \qquad (1.15)$$

where $\left[A\overset{\otimes}{,}B\right]_- = A\otimes B - B\otimes A$.

Still ignoring impurity scattering,

$$\Sigma_n\left(\vec{r},t\right)=\left(\Sigma_n\left(\vec{r},t\right)\right)_{pairing}=\begin{pmatrix}\hat{\Delta}_n\left(\vec{r},t\right) & 0\\ 0 & \hat{\Delta}_n\left(\vec{r},t\right)\end{pmatrix},\hat{\Delta}_n\left(\vec{r},t\right)=\begin{pmatrix}0 & \Delta_n\left(\vec{r},t\right)\\ -\Delta_n^*\left(\vec{r},t\right) & 0\end{pmatrix}$$

(1.16)

The pairing potential is defined as

$$\Delta_n\left(\vec{r}\right)=-\sum_m V_{nm}\left\langle\hat{\psi}_{m\downarrow}(\vec{r})\hat{\psi}_{m\uparrow}(\vec{r})\right\rangle=\frac{i}{2}\sum_m V_{nm}\lim_{(\vec{r}',t')\to(\vec{r},t)}Tr\left[\left(\tau^1-i\tau^2\right)\hat{G}_n^K\left(\vec{r},t;\vec{r}',t'\right)\right],$$
$$\tau^1=\begin{pmatrix}0 & 1\\ 1 & 0\end{pmatrix},\tau^2=\begin{pmatrix}0 & -i\\ i & 0\end{pmatrix} \qquad (1.17)$$

## 2. The Eilenberg equations

Eilenberg derived transportlike equations for one-band superconductor [51,49,50].We would like to obtain transportlike equations for a two-band superconductor. To derive the quantum kinetic equation we introduce the mixed (Wigner) coordinates

$$\vec{R} = \frac{\vec{r} + \vec{r}'}{2}, \vec{\rho} = \vec{r} - \vec{r}',$$
$$T = \frac{t + t'}{2}, \vartheta = t - t', \qquad (2.1)$$
$$A_{ij}\left(\vec{r}, t; \vec{r}', t'\right) = A_{ij}\left(\vec{R} + \frac{\vec{\rho}}{2}, T + \frac{\vartheta}{2}; \vec{R} - \frac{\vec{\rho}}{2}, T - \frac{\vartheta}{2}\right) \equiv A_{ij}\left(\vec{R}, T; \vec{\rho}, \vartheta\right)$$

and then define a Fourier transform

$$A_{ij}\left(\vec{R}, T; \vec{\rho}, \vartheta\right) = \int d\vec{p} \int dE e^{-iE\vartheta} e^{i\vec{p}\vec{\rho}} A_{ij}\left(\vec{R}, T; \vec{p}, E\right) \qquad (2.2)$$

If the derivatives are small, we need to take only the first order approximation of the commutator

$$\left[A \overset{\otimes}{,} B\right]_{-} \left(\vec{R}, T; \vec{p}, E\right) = (A \otimes B - B \otimes A)\left(\vec{R}, T; \vec{p}, E\right) = \left[A, B\right]_{-} +$$
$$+ \frac{i}{2}\left[\left(\left\{\partial_T A, \partial_E B\right\} - \left\{\partial_E A, \partial_T B\right\}\right) + \left(\left\{\partial_{\vec{R}} A, \partial_{\vec{p}} B\right\} - \left\{\partial_{\vec{p}} A, \partial_{\vec{R}} B\right\}\right)\right]_{-} \qquad (2.3)$$

where $\{A, B\}$ stands for the anticommutator.

Taking into account impurity scattering, the Dyson equation (1.15) can be written as

$$\left[i\tau^{(3)}\frac{\partial}{\partial t} + \frac{1}{2m_n}\left(\frac{\partial}{\partial \vec{r}} - ie\tau^{(3)}\vec{A}(\vec{r}, t)\right)^2 - e\varphi(\vec{r}, t) + \mu - \zeta_n - (\Sigma_n)_{pairing} - (\Sigma_n)_{imp} \overset{\otimes}{,} G_n\right]_{-} = 0 \qquad (2.4)$$

where the impurity self-energy $(\Sigma_n)_{imp}$ includes contributions from both spin-flip and spin-independent elastic scattering

$$(\Sigma_n)_{imp} = (\Sigma_n)_0 + (\Sigma_n)_{sf}$$

Without external fields equation (2.1) can be written as

$$\left[ i\tau^{(3)} \frac{\partial}{\partial t} + \frac{1}{2m_n} \frac{\partial^2}{\partial \vec{r}^2} + \mu - \zeta_n - \left(\Sigma_n\right)_{pairing} - \left(\Sigma_n\right)_{imp} \overset{\otimes}{,} G_n \right]_- = 0 \tag{2.5}$$

The elastic contribution to the self-energy can be written as

$$\left(\Sigma_n\right)_0 \left(\vec{p}\right) = N_{imp} \sum_{m=1,2} \int d\vec{p}' \left| u_{nm} \left(\vec{p} - \vec{p}'\right)\right|^2 G_m(\vec{p}) \tag{2.6}$$

Here $u_{nm}$ is the matrix element of impurity potential, and $N_{imp}$ the number of impurities per unit volume. We assume that $u(\vec{p})$ is independent of the magnitude of $\vec{p}$, so that

$$\left(\Sigma_n\right)_0 \left(\vec{p}\right) = N_{imp} N_0 \sum_{m=1,2} \int d\varepsilon_{\vec{p}'} \frac{d\Omega_{\vec{p}'}}{4\pi} \left| u_{nm} \left(\hat{p} \cdot \hat{p}'\right)\right|^2 G_m(\vec{p}) \tag{2.7}$$

Where $N_0$ is the density of states at the Fermi energy, and $d\Omega_{\vec{p}}$ is an element of solid angle in momentum space, $\hat{p}$ is the unity vector in the direction of $\vec{p}$. Defining the elastic scattering rate $\frac{1}{\tau_{nm}}$ in the Born approximation by

$$\frac{1}{\tau_{nm}} = 2\pi N_{imp} N_0 \int \frac{d\Omega_{\vec{p}'}}{4\pi} \left| u_{nm} \left(\hat{p} \cdot \hat{p}'\right)\right|^2 \tag{2.8}$$

we can write

$$\left(\Sigma_n\right)_0 \left(\vec{p}\right) = \sum_{m=1,2} \frac{1}{2\pi\tau_{nm}} \int d\varepsilon_{\vec{p}} G_m(\vec{p}) \tag{2.9}$$

Similarly, for the contribution from spin-flip scattering due to magnetic impurities one obtains

$$\left(\Sigma_n\right)_{sf}\left(\vec{p}\right)=\sum_{m=1,2}\frac{1}{2\pi(\tau_{nm})_{sf}}\int d\varepsilon_{\vec{p}}\tau^{(3)}G_m(\vec{p})\tau^{(3)} \tag{2.10}$$

where

$$\frac{1}{(\tau_{nm})_{sf}}=2\pi N_{magn.imp}N_0 s(s+1)\int\frac{d\Omega_{\vec{p}'}}{4\pi}\left|u_{nm}\left(\hat{p}\cdot\hat{p}'\right)\right|^2 \tag{2.11}$$

$N_{magn.imp}$ is the number of magnetic impurities per unit volume, $s$ is the spin of the impurity, $\frac{1}{\tau_{nn}}$ is the intraband scattering rate within the n-th band, $\frac{1}{\tau_{nm}}(n\neq m)$ is the interband scattering rate. Converting the left hand side (2.5) to relative coordinates, and using (2.3), we obtain

$$\left[\tau^{(3)}E+\left(\Sigma_n\right)_{pairing},G_n\right]_- +i\vec{v}_{Fn}\cdot\partial_{\vec{R}}G_n-\left[\left(\Sigma_n\right)_0+\left(\Sigma_n\right)_{sf},G_n\right]_- =0 \tag{2.12}$$

where $\vec{v}_{Fn}$ is the Fermi velocity in the n-th band. We define the quasiclassical Green's functions

$$g_n\left(\vec{R},\hat{p},t,t'\right)=\frac{i}{\pi}\int d\varepsilon_{\vec{p}}G_n\left(\vec{R},\vec{p},t,t'\right) \tag{2.13}$$

Written in terms of quasiclassical Green's functions the equation (2.12) becomes

$$\left[\tau^{(3)}E+\left(\Sigma_n\right)_{pairing},g_n\right]_- +i\vec{v}_{Fn}\cdot\partial_{\vec{R}}g_n-\left[\left(\sigma_n\right)_0+\left(\sigma_n\right)_{sf},g_n\right]_- =0 \tag{2.14}$$

where

$$\left(\sigma_n\right)_0=-\sum_{m=1,2}\frac{i}{2\tau_{mn}}(g_m)_{aver} \tag{2.15}$$

and

$$\left(\sigma_n\right)_{sf} = -\sum_{m=1,2} \frac{i}{2\left(\tau_{mn}\right)_{sf}} \tau^{(3)} (g_m)_{aver} \tau^{(3)} \tag{2.16}$$

The subscript "aver" to the quasiclassical Green's functions denotes that they are averaged over all angles of momentum. Equation (2.14) is Eilenberg's equation for two-band superconductors.

## 3. The Usadel equation

For a dirty superconductor, a short mean-free path makes the theory simple, since the motion of the electrons is always nearly isotropic [52, 49, 50].

Using $G_n = \begin{pmatrix} \hat{G}_n^R & \hat{G}_n^K \\ 0 & \hat{G}_n^A \end{pmatrix}$ and (2.12) we get

$$g_n = \begin{pmatrix} \hat{g}_n^R & \hat{g}_n^K \\ 0 & \hat{g}_n^A \end{pmatrix} \tag{3.1}$$

where $\hat{g}_n^R, \hat{g}_n^A, \hat{g}_n^K$ are $2\times 2$ matrices. Normalization condition for the quasiclassical Green's functions is

$$g_n g_n = \tau^{(0)},$$
$$\tau^{(0)} = \begin{pmatrix} \tau^0 & 0 \\ 0 & \tau^0 \end{pmatrix}, \tau^0 = \begin{pmatrix} 1 & 0 \\ 0 & 1 \end{pmatrix} \tag{3.2}$$

Equation (3.2) is equivalent to the three $2\times 2$ matrix equations

$$\hat{g}_n^R \hat{g}_n^R = \tau^0 \tag{3.3}$$

$$\hat{g}_n^A \hat{g}_n^A = \tau^0 \tag{3.4}$$

$$\hat{g}_n^R \hat{g}_n^K + \hat{g}_n^K \hat{g}_n^A = 0 \tag{3.5}$$

From Eilenberg's equation (2.14) we obtain

$$\left[E\tau^3+\hat{\Delta}_n,\hat{g}_n^R\right]_- + i\vec{v}_{Fn}\cdot\partial_{\vec{R}}\hat{g}_n^R - \left[\hat{\sigma}_n^R,\hat{g}_n^R\right]_- = 0 \tag{3.6}$$

$$\left[E\tau^3+\hat{\Delta}_n,\hat{g}_n^A\right]_- + i\vec{v}_{Fn}\cdot\partial_{\vec{R}}\hat{g}_n^A - \left[\hat{\sigma}_n^A,\hat{g}_n^A\right]_- = 0 \tag{3.7}$$

$$\left[E\tau^3+\hat{\Delta}_n,\hat{g}_n^K\right]_- + i\vec{v}_{Fn}\cdot\partial_{\vec{R}}\hat{g}_n^K - \hat{\sigma}_n^R\hat{g}_n^K - \hat{\sigma}_n^K\hat{g}_n^A + \hat{g}_n^R\hat{\sigma}_n^K + \hat{g}_n^K\hat{\sigma}_n^A = 0 \tag{3.8}$$

where $\sigma_n = (\sigma_n)_0 + (\sigma_n)_{sf} = \begin{pmatrix} \hat{\sigma}_n^R & \hat{\sigma}_n^K \\ 0 & \hat{\sigma}_n^A \end{pmatrix}$. Here $\hat{\sigma}_n^R,\hat{\sigma}_n^A,\hat{\sigma}_n^K$ are $2\times2$ matrices. If elastic impurity scattering is strong the motion of electrons is not ballistic, but diffusive. It makes sense to average the self-energy and Green's functions over the directions of the momentum. In the dirty limit, the quasiclassical Green's function will be almost isotropic , and an expansion in spherical harmonics needs only keep the s- and p-wave parts

$$\begin{aligned} g_n &= g_{ns} + \hat{p}g_{np}, \\ \sigma_n &= (\sigma_n)_0 + (\sigma_n)_{sf} = \sigma_{ns} + \hat{p}\sigma_{np} \end{aligned} \tag{3.9}$$

The assumption is that $g_{np} \ll g_{ns}, \sigma_{np} \ll \sigma_{ns}$ .Substituting eq. (3.9) into the (2.15), (2.16), we obtain

$$\sigma_{ns} = -\sum_{m=1,2}\frac{i}{2\tau_{mn}}g_{ms} - \sum_{m=1,2}\frac{i}{2\left(\tau_{mn}\right)_{sf}}\tau^{(3)}g_{ms}\tau^{(3)} \tag{3.10}$$

$$\sigma_{np} = -\frac{i}{2}\sum_{m=1,2}\left[\frac{1}{\tau_{mn}} - \frac{1}{\left(\tau_{mn}\right)_{tr}}\right]g_{mp} \tag{3.11}$$

where $(\tau_{nm})_{tr}$ is the impurity transport relaxation time

$$\frac{1}{(\tau_{nm})_{tr}} = 2\pi N_{imp} N_0 \int \frac{d\Omega_{\vec{p}'}}{4\pi} \left| u_{nm}\left(\hat{p}\cdot\hat{p}'\right)\right|^2 \left(1-\hat{p}\cdot\hat{p}'\right) \quad (3.12)$$

Inserting (3.9) into Eilenberg's equation we obtain

$$\begin{aligned}&\left[\tau^{(3)}E+\left(\Sigma_n\right)_{pairing}, g_{ns}\right]_- + \hat{p}\left[\tau^{(3)}E+\left(\Sigma_n\right)_{pairing}, g_{np}\right]_- + iv_{Fn}\hat{p}\cdot\partial_{\vec{R}}(g_{ns}+\hat{p}g_{np})\\ &-\left[\sigma_{ns}, g_{ns}\right]_- - \hat{p}\left[\sigma_{ns}, g_{np}\right]_- - \left[\sigma_{np}, g_{ns}\right]_- - \hat{p}\cdot\hat{p}\left[\sigma_{np}, g_{np}\right]_- = 0\end{aligned} \quad (3.13)$$

Equation (3.13) for the dirty limit can be split into even and odd parts with respect to momentum

$$\begin{aligned}&\left[\tau^{(3)}E+\left(\Sigma_n\right)_{pairing}, g_{ns}\right]_- + iv_{Fn}(\hat{p}\cdot\hat{p})\partial_{\vec{R}} g_{np}\\ &+\sum_{\substack{m=1,2\\(m\neq n)}} \frac{i}{2\tau_{mn}}\left[g_{ms}, g_{ns}\right]_- + \sum_{m=1,2}\frac{i}{2\left(\tau_{mn}\right)_{sf}}\left[\tau^{(3)}g_{ms}\tau^{(3)}, g_{ns}\right]_- = 0\end{aligned} \quad (3.14)$$

and

$$\left[\tau^{(3)}E+\left(\Sigma_n\right)_{pairing}, g_{np}\right]_- + iv_{Fn}\partial_{\vec{R}} g_{ns} - \frac{i}{2}\sum_{m=1,2}\frac{1}{\left(\tau_{mn}\right)_{tr}}\left[g_{mp}, g_{ns}\right]_- = 0 \quad (3.15)$$

Averaging eq. (3.14) over all directions of momentum gives

$$\begin{aligned}&\left[\tau^{(3)}E+\left(\Sigma_n\right)_{pairing}, g_{ns}\right]_- + \frac{i}{3}v_{Fn}\partial_{\vec{R}} g_{np}\\ &+\sum_{\substack{m=1,2\\(m\neq n)}} \frac{i}{2\tau_{mn}}\left[g_{ms}, g_{ns}\right]_- + \sum_{m=1,2}\frac{i}{2\left(\tau_{mn}\right)_{sf}}\left[\tau^{(3)}g_{ms}\tau^{(3)}, g_{ns}\right]_- = 0\end{aligned} \quad (3.16)$$

If elastic scattering is strong, the energy of impurity scattering is much larger than any energy in the problem so that the first term in the equation (3.15) can be neglected compared to the third and we obtain

$$iv_{Fn}\partial_{\vec{R}} g_{ns} - \frac{i}{2}\sum_{m=1,2}\frac{1}{\left(\tau_{mn}\right)_{tr}}\left[g_{mp}, g_{ns}\right]_- = 0 \quad (3.17)$$

Equation (3.17) can be rewriten in the following form

$$iv_{Fn}\partial_{\vec{R}}g_{ns}-\frac{i}{2}\frac{1}{\left(\tau_{nn}\right)_{tr}}\left[g_{np},g_{ns}\right]_{-}-\frac{i}{2}\sum_{\substack{m=1,2\\(m\neq n)}}\frac{1}{\left(\tau_{mn}\right)_{tr}}\left[g_{mp},g_{ns}\right]_{-}=0 \tag{3.18}$$

From the normalization condition (3.2) we also obtain

$$\begin{aligned} g_{ns}g_{ns} &= \tau^{(0)}, \\ g_{ns}g_{np}+g_{np}g_{ns} &= 0 \end{aligned} \tag{3.19}$$

Upon inserting into (3.18), we get

$$il_n\partial_{\vec{R}}g_{ns}+ig_{ns}g_{np}-\frac{i}{2}\sum_{\substack{m=1,2\\(m\neq n)}}\frac{\left(\tau_{nn}\right)_{tr}}{\left(\tau_{mn}\right)_{tr}}\left[g_{mp},g_{ns}\right]_{-}=0 \tag{3.20}$$

where $l_n=v_{Fn}\left(\tau_{nn}\right)_{tr}$ is the impurity mean free path in the n-th band. Within the model of weak interband impurity scattering $\left(\tau_{nn}\right)_{tr}\ll\left(\tau_{mn}\right)_{tr}$, Eq. (3.20) takes the simple form $g_{ns}g_{np}=-l_n\partial_{\vec{R}}g_{ns}$. Multiplying this equation by $g_{ns}$ on the left and using Eq. (3.19) we obtain

$$g_{np}=-l_ng_{ns}\partial_{\vec{R}}g_{ns} \tag{3.21}$$

Putting this into Eq. (3.16) we obtain

$$\begin{aligned} &\left[\tau^{(3)}E+\left(\Sigma_n\right)_{pairing},g_{ns}\right]_{-}-iD_n\partial_{\vec{R}}\left(g_{ns}\partial_{\vec{R}}g_{ns}\right) \\ &+\sum_{\substack{m=1,2\\(m\neq n)}}\frac{i}{2\tau_{mn}}\left[g_{ms},g_{ns}\right]_{-}+\sum_{m=1,2}\frac{i}{2\left(\tau_{mn}\right)_{sf}}\left[\tau^{(3)}g_{ms}\tau^{(3)},g_{ns}\right]_{-}=0 \end{aligned} \tag{3.22}$$

where $D_n=(1/3)l_nv_{Fn}$ is the diffusion coefficient. The Eq. (3.22) is the system of coupled Usadel's equations in which all microscopic details are hidden in the diffusion coefficients for each Fermi surface sheet and the interband scattering rates $1/\tau_{mn}(m\neq n)$ . Ignoring spin-flip scattering, eq. (3.22) can be written

$$\left[\tau^{(3)}E+\left(\Sigma_n\right)_{pairing},g_{ns}\right]_- -iD_n\partial_{\vec{R}}\left(g_{ns}\partial_{\vec{R}}g_{ns}\right)+\sum_{\substack{m=1,2\\(m\neq n)}}\frac{i}{2\tau_{mn}}\left[g_{ms},g_{ns}\right]_-=0 \qquad (3.23)$$

Representing $g_{ns}$ as a matrix

$$g_{ns}=\begin{pmatrix}\hat{g}^R_{ns} & \hat{g}^K_{ns}\\ 0 & \hat{g}^A_{ns}\end{pmatrix}, \qquad (3.24)$$

Eq. (3.23) becomes

$$\left[E\tau^3+\hat{\Delta}_n,\hat{g}^R_{ns}\right]_- -iD_n\partial_{\vec{R}}\left(\hat{g}^R_{ns}\partial_{\vec{R}}\hat{g}^R_{ns}\right)=0 \qquad (3.25)$$

$$\left[E\tau^3+\hat{\Delta}_n,\hat{g}^A_{ns}\right]_- -iD_n\partial_{\vec{R}}\left(\hat{g}^A_{ns}\partial_{\vec{R}}\hat{g}^A_{ns}\right)=0 \qquad (3.26)$$

$$\begin{aligned}&\left[E\tau^3+\hat{\Delta}_n,\hat{g}^K_{ns}\right]_- -iD_n\partial_{\vec{R}}\left(\hat{g}^R_{ns}\partial_{\vec{R}}\hat{g}^K_{ns}+\hat{g}^K_{ns}\partial_{\vec{R}}\hat{g}^A_{ns}\right)\\&+\sum_{\substack{m=1,2\\(m\neq n)}}\frac{i}{2\tau_{mn}}\left(\hat{g}^R_{ms}\hat{g}^K_{ns}+\hat{g}^K_{ms}\hat{g}^A_{ns}-\hat{g}^R_{ns}\hat{g}^K_{ms}-\hat{g}^K_{ns}\hat{g}^A_{ms}\right)=0\end{aligned} \qquad (3.27)$$

The system of equations (3.25)-(3.27) is the "dirty limit" of the system (3.6)-(3.8).This system of differential equations must be supplemented by a proper boundary conditions. For unbounded superconductor one has match the equilibrium solution at large distances. To solve the equations for bounded superconductor one needs some boundary conditions at surfaces and interfaces which are in general partially transparent. The boundary transparency of quasiparticles from different bands is important parameter. An extended discussion of the reflection and transmission processes at surfaces of two-band superconductor will be given elsewhere.

In conclusion, in this paper we developed kinetic equations for two-band dirty superconductor using the real-time technique.

We would like to thank Prof. J.Berger for his support of the present work. This research was supported by the Israel Science Foundation, grant No\249/10.